\documentclass[12pt]{JHEP3}
\usepackage{amsmath}
\usepackage{amsfonts}
\usepackage{amssymb}
\usepackage{epsfig}
\def\beq{\begin{equation}}
\def\beql#1{\begin{equation}\label{eq:#1}}
\def\eeq{\end{equation}}

\newcommand{\tr}{{\rm tr}~ }
\newcommand{\comment}[1]{}

\newcommand{\pasl}{\pa\kern-.55em /}

\newcommand{\ksl}{k\kern-.55em /}

\DeclareFixedFont{\xiiss}{OT1}{cmss}{m}{n}{12}
\DeclareFixedFont{\ixss}{OT1}{cmss}{m}{n}{9}
\DeclareFixedFont{\cmrnine}{OT1}{cmr}{m}{n}{9}
\newcommand{\field}[1]{\mathbb{#1}}
\newcommand{\BC}{{\field C}}

\newcommand{\BZ}{{\field Z}}

\newcommand{\CCs}{\hbox{\ixss C\kern-.4emI}}
\newcommand{\ZZs}{\hbox{\ixss Z\kern-.4emZ}}

\newcommand{\littlefig}[2]{
 \epsfxsize=#2in
 \epsfbox{#1}
}

\title{D-brane realizations of runaway behavior and 
moduli stabilization}

\author{David Berenstein \thanks{dberens@ias.edu}\\
 School of Natural Sciences,
Institute for Advanced Study, Princeton, NJ 08540, USA }
\abstract{ In this paper we find examples of moduli stabilization 
and runaway behavior which can be treated exactly. 
This is shown for supersymmetric field theories which can be realized on 
the world volume of 
D-branes. From a geometric point of view, these field theories
lift moduli spaces of vacua by deforming lines of  
singularities where supersymmetric fractional branes can be 
located in the geometry without D-branes.}

\keywords{Supersymmetric gauge theories, moduli stabilization}
\begin{document}

\section{Introduction}

Supersymmetric string compatifications on Calabi-Yau manifolds 
are usually characterized 
by having continuous families of solutions that satisfy the string equations
of motion, this family is called the moduli space of a compactification.
One can move inside the family by exciting closed string 
fields, and since the total change of energy is zero, these fields are 
massless. These massless fields are scalars with respect to the four 
dimensional physics, and are called moduli.

It is believed that non-perturbative effects can cause an effective 
superpotential on the moduli space which lifts the degeneracy of these vacua
\cite{Wnp}, 
and leaves behind a finite number of supersymmetric vacua. 
Thus there
is no true moduli space, but only an asymptotic region where some
of the moduli fields can be considered to be very light. This is usually a region with
 runaway behavior, and moduli roll towards
ten dimensional flat space.

Most of the full structure of moduli space is inaccessible to 
computations because the string dilaton is one of the moduli, and 
we have very little 
understanding of the theory at strong coupling to determine the 
structure 
of the moduli space and the superpotential on it. 

In most circumstances all we understand is an expansion of the theory about 
a weak coupling point, and we are forced to look for solutions which do 
not stray too far from the weak coupling regime.

For most results however, one can not sum the full set of
non-perturbative corrections, and the effective superpotential on the moduli 
space is given roughly by
\begin{equation} 
W_{eff} = W_1+W_2+\hbox{Other uncontrolled non-perturbative corrections}
\end{equation}
where one believes $W_1$ and $W_2$ dominate in some region of moduli 
space. Here, we explicitly write two contributions that are associated with 
distinct dependence on the closed string dilaton, so that one can
 balance the two effects and produce a finite vev for the dilaton,
 hopefully  in a perturbative regime for the calculation of some 
quantities 
(this has been called a racetrack scheme. It was discussed originally
in \cite{Kras}. For a 
more recent discussion
see \cite{DS}).

In this paper we will explore toy models for
 moduli stabilization in supersymmetric 
field theories. The main points of the paper are to exploit the recent 
advances in describing the structure of supersymmetric vacua by 
matrix models \cite{DV}, and to geometrize the 
field theory behavior into aspects of the geometry 
of a system of D-branes, so that we can come into contact with the 
stabilization of moduli for more geometric setups. At the same time, 
retaining just a field theory calculation and decoupling gravity 
and the dilaton, because we are taking a non-compact Calabi-Yau 
geometry.

The main advantage of the setup described in this paper is that it can be 
argued to be exact, due to their 
relations to matrix models. In this sense it is now 
possible to make certain arguments on the whole moduli space of a 
theory, 
instead of a more
usual procedure of taking limits in various regions where different
 manipulations give a tractable answer \cite{IT}.

This program should be viewed as baby steps towards producing vacua 
as described in the work of Kachru et al. \cite{KKLT}, 
where first one describes a supersymmetric 
compactification, and at the very end one adds anti-D3 branes to break 
supersymmetry on an F-theory 
geometry. this finla step produces
a De-Sitter like vacuum in string theory. It has been 
argued by Susskind \cite{S} based on ideas by Bousso and Polchinski
\cite{BP}
that there is possibly a very large number  of these models.   
Under these circumstances 
it is important to understand under what conditions can
one trust the calculations that one is performing. See also the recent 
discussion by Douglas \cite{Dou}, where an attempt is made to count vacua.

The paper is organized as follows:

In section \ref{sec:general} we study the topology of moduli spaces 
and the conditions under which classical moduli spaces can be lifted by 
quantum corrections. We argue that there need to be singularities in 
codimension one on the classical moduli space for this to happen.
In section \ref{sec:ADS} we give a D-brane realization 
of the Affleck-Dine-Seiberg system by putting a collection of fractional 
branes on a $\BC^3/\BZ_2\times\BZ_2$ singularity. We study the geometry of 
the system in detail and show that confining fractional branes remove the 
three lines of singularities when one computes the deformed geometry. We also 
study a Seiberg-dual version of the system which allows for 
more easy generalizations. Next, in section \ref{sec:stab}, we study a 
variation of a racetrack scheme which allows for gaugino condensation in two 
gauge groups to stabilize the position of a brane.
This example can be obtained by deformations of an ${\cal N}=2$
theory softly broken to ${\cal N}= 1$. The theory
has various vacua with very different properties. We give a 
qualitative analysis of the light spectrum of particles in some of the
vacua which are interesting. We close the paper with some concluding remarks.

\section{Lifting moduli spaces}\label{sec:general}

Given a classical moduli space of vacua, we can ask what properties of the 
moduli space are necessary to have a superpotential generated by 
quantum effects on the classical moduli space. The basic property of the 
effective superpotential is that it is given by a holomorphic 
(complex analytic) 
function on the moduli space of vacua. Traditional setups include a
conserved $R$-charge which makes it possible to argue for the exact form 
of the superpotential. A review with many examples and guide to the literature
can be found in \cite{SS}. The new matrix model 
ideas \cite{DV,DGLVZ,CDSW} can be argued to be 
exact, irrespective of the presence of these additional symmetries,
and therefore one can now study many examples which were not possible in 
the past.

A very important point to  remember is that the moduli 
spaces given by field theories are usually noncompact, with 
infinity being given by the region of large vevs for some fields in 
the SUSY field theory. Under good conditions, the infinity will be 
weakly coupled and therefore quantum corrections will be small.
In effect, this gives us a compactification of the moduli space of vacua, and 
then the superpotential will be a complex analytic 
function (it could be multi-valued)  
on the compactified moduli space. If this function is non-constant, 
then because it is holomorphic it will necessarily have singularities 
somewhere in the middle of the moduli space. 
These are 
either monodromies or poles 
and should be associated to some massless particle being present at 
the singularity. Infinity can also have monodromies associated to it, so 
if one knows the structure of the singularities it is possible to guess 
the superpotential function inside the moduli space, by requiring a 
fixed type of behavior at each singularity.
Necessarily all of  these singularities are of complex 
codimension one in the moduli space.
It is exactly this style of reasoning that produced the solution of 
${\cal N}=2$ field theories by Seiberg and Witten \cite{SW}, except that 
the holomorphic object was the infrared 
gauge coupling on the moduli space, and the holomorphic map
was to the upper half plane, 
and then modded out by the $SL(2,Z)$ S-duality group. 

 Indeed, these singularities associated to massless particles 
should be already 
present in the classical theory, so we find that the geometry of the moduli 
space requires classical singularities in codimension one. 
If these are 
not present, then the moduli space is not lifted, and the 
only other possibility for quantum effects on the moduli space 
is that it becomes deformed.

The analysis above can be done branch by branch on the moduli space, so it is
possible to have theories where some branches of the moduli space
are lifted and some others are not.

Obviously the above arguments can be further clarified with the help
of some examples.

In the case of ${\cal N} = 4$ SYM, for gauge group $U(N)$ the classical
moduli space 
is given by $\BC^3/S_N$, the symmetric product of $N$ copies of $\BC^3$, and 
it is described by a set of three commuting matrices of rank $N$, which 
can be diagonalized. The singularities of the moduli space occur at places 
where there is enhanced gauge symmetry. This is a set that requires
 us to fix 
three pairs of eigenvalues simultaneously. This phenomenon
occurs in complex codimension three, and therefore the 
moduli space is not lifted by quantum corrections. Indeed, from the 
high amount of symmetry the moduli space is not deformed at all.

A second example consists of a field theory one of whose branches of 
 moduli space is a 
(non-compact) Calabi-Yau geometry. In this case, the Calabi-Yau 
geometry can only have singularities in codimension two or higher,
 so again, the moduli space can not be lifted by quantum corrections.
This example is relevant for a probe brane in the conifold 
geometry with fractional branes places at the conifold 
(the Klebanov-Strassler system \cite{KS}). 
In this case the geometry gets deformed.

Finally, we can consider the Affleck-Dine-Seiberg \cite{ADS} 
field theory with gauge 
group $SU(N)$ and $N_F<N$ quarks $Q_i, \tilde Q_i$. One can argue that 
at generic points in the classical moduli space that the theory has an 
unbroken $SU(N-N_F)$ gauge group, which has a gaugino condensate.

The moduli space is parametrized by the $N_F\times N_F$ meson matrix 
$M_{ij}=Q_i\tilde Q_j$. Generic points in moduli space are characterized by 
$M$ having maximal rank 
$N_F$. The singularities in the moduli space are characterized by $M$ having 
smaller rank. The order parameter that determines this property 
is wether the single equation on 
moduli space $\det(M)=0$ is true or not.
 Thus, these singularities occur in codimension one, 
and are associated to the field theory having a point of enhanced symmetry 
$SU(N-N_F+1)$. In this example, the classical moduli space has an effective 
superpotential given by \cite{ADS}
\begin{equation}
W_{eff} \sim \left[\frac{\Lambda^{3N-N_f}}{\det M}\right]^{(N-N_f)^{-1}}
\end{equation}
 Obviously this effective 
superpotential is singular exactly at the classical singularities in moduli 
space. Here one does not get a pole at the singularities 
unless $N_f=N-1$. This is the same 
condition required for the superpotential to be generated by 
instantons. In the other cases there are monodromies at the singularities, 
which can be associated to motion between the $(N-N_F)$ vacua of the
pure $SU(N-N_F)$ theory.

\section{Affleck-Dine-Seiberg with D-branes.}\label{sec:ADS}

Now we want to use the results of the past section to start building 
D-brane field theories which have runaway behavior or moduli stabilization.

As we saw, we require that the moduli space have singularities in codimension 
one. The simplest such moduli space would be a one complex dimensional 
manifold. A D-brane with such a moduli space is usually a fractional 
brane at a curve of singularities, e.g. a D5 brane wrapped on a
holomorphic two cycle which has shrunk to zero size.

The natural place to find such geometries is in orbifold with fixed lines
of singularities. For example, let us take take $\BC^3/\BZ_2$. Here, the 
fixed point set of the group action is the singular locus, and it gives 
rise to an ${\cal N}=2$ SYM theory for a fractional brane. However, this 
theory has too much supersymmetry, and there are no singularities in the 
moduli space of a single D-brane. To remedy this situation, we can introduce
a marked point on the moduli space, by performing an additional orbifold, to 
obtain the $\BC^3/\BZ_2\times\BZ_2$ orbifold.

This theory is described by the following quiver diagram

\littlefig{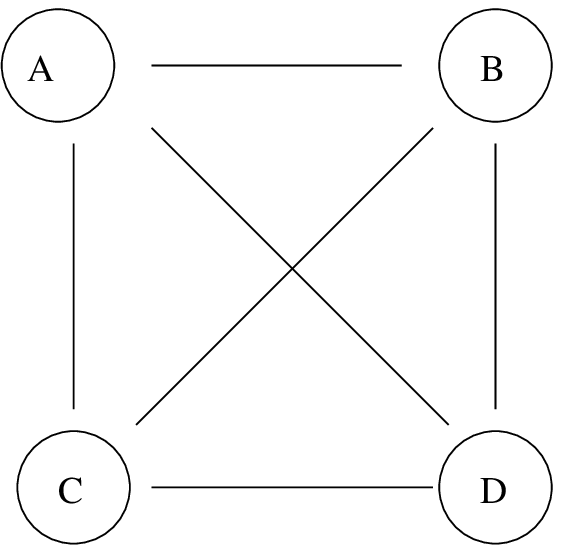}{2}

where we have labeled the nodes of the quiver with capital letters 
$A,B,C,D$. The quiver is not-chiral, and all edges correspond to two 
chiral multiplets with opposite quantum numbers under the gauge group.
We will label these as $\phi_{XY}$ where $X$ and $Y$ indicate the two 
gauge groups under which it is charged.

The geometry of the orbifold is given by one equation in four 
variables
\begin{equation}
uvw=t^2\label{eq:z2z2orb}
\end{equation}
and it contains three lines of singularities meeting at the origin. These 
lines of singularities correspond to the locus $u,v=0$, $v,w=0$ and $w,u=0$.
A single brane in the bulk has brane content $A+B+C+D$, and for this brane the 
variables $u,v,w,t$ can be identified as follows
\begin{equation}
u = \phi_{AB}\phi_{BA}, v =\phi_{AC}\phi_{CA}, w=\phi_{AD}\phi_{DA},
t=\phi_{AB}\phi_{BD}\phi_{DA}
\end{equation}
A straightforward manipulation of the F-terms show that these variables
 satisfy equation \ref{eq:z2z2orb}.

Now, the fractional branes at the singularities are constructed from 
combinations of two different fractional branes like $A+B$. This brane 
has a one dimensional moduli space characterized by the vev of the gauge 
invariant field $\phi_{AB}\phi_{BA}=u$ which gives us a brane which spans
the $u$-line of singularities. Notice that we have a marked point at the 
origin where the gauge group is enhanced to $U(1)\times U(1)$.

This configuration does not get it's moduli space lifted however, 
since it can be argued to be a configuration which can be obtained by
orbifolding an ${\cal N}=2$ theory without adding extra ${\cal N}=1$ 
fractional branes.

However, we can consider the following configuration of branes
$N A + N_FB$. This configuration for $N_A\neq N_B$ confines either the gauge 
group $U(N_A)$ or the gauge group $U(N_B)$, and is exactly the gauge theory 
that one would obtain from the Affleck-Dine-Seiberg system if one gauged the 
vector like $U(N_F)$ flavor symmetry. We will now take $N>N_F$, and we know 
that this particular configuration has runaway behavior.

Let us consider a generic point in the moduli space. This will be 
characterized by the meson matrix $M=\phi_{BA}\phi_{AB}$. With the 
$U(N_F)$ symmetry we can diagonalize it, and we obtain $N_F$ branes at generic
points in  the $u$ 
curve of singularities. Also, we get $N-N_F$ branes stuck at the origin.

These branes at the origin in the low energy effective field theory are pure 
$U(N-N_F)$ and the $SU(N-N_F)$ confines. 

Confinement in geometric situations usually leads to a geometric transition: 
a deformation of the complex structure due to exchanging even cycles where 
branes can wrap by fluxes \cite{GV,KS}. 
The shape of the deformation in this case can be argued by 
holomorphy \cite{Sexact,Shol,Snal}.

In the field theory  there is a non-anomalous $U(1)^3$ 
symmetry which is the remnant of the R-symmetry of the unorbifolded gauge 
theory. $u,v,w$ are charged under these global symmetries and one can not
deform the cubic term in the equation without breaking these symmetries. 
The only allowed deformation can be put in the form
\begin{equation}\label{eq:z2z2def}
uvw = t^2 + c
\end{equation}
where $c$ is a constant. We will leave a matrix model derivation of this 
effect for the appendix.

Given that this is the form of the deformation, we can readily understand why 
the moduli space for fractional branes is lifted. Clearly the above 
geometry is regular for $c\neq 0$, all of the fractional branes get their 
moduli space lifted because there are no singularities left over where we 
can support a supersymmetric fractional brane. Again, the result of the 
appendix shows that the moduli space of a brane in the bulk is not 
lifted.

Also, one can derive the Affleck-Dine-Seiberg superpotential directly from 
matrix models (see for example \cite{DJ,BR,Ohta}), and the result is given 
exactly by 
\begin{equation}\label{eq:su1}
W_{eff} = (N-N_F) (S\log S-S)+\tau S + S \log\det(M)
\end{equation}
where $S$ is the gaugino condensate for the unbroken gauge group.
So long as $S\neq 0$, one can see that there is no saddle point for $M$. 

Surprisingly, this has implication for the conformal field theory 
associated to the singularity. One can consider points in the moduli 
space where the branes are split according to configurations
\begin{equation}
2N A + N(B+C) + N(C+D) +N(B+D)
\end{equation}
In these configurations at generic points in moduli space the gauge group 
$U(2N)_A$ reduces to pure gauge theory in the deep infrared, and it confines.
The geometric transition described before still takes place, and the branch
of moduli space is lifted. In essence one can show that any branch which
leaves confining branes at the origin is lifted.

This is a slightly surprising result, as we are used to thinking of 
orbifolds of ${\cal N}=4$ gauge theory as being essentially classical
objects, and the classical moduli space as being exact. This is only
true when the  moduli correspond exclusively to
branes in the bulk (the discussion of codimension of singularities in the 
previous section gives that result).

Now, let us explore a related theory where we take a formal 
Seiberg duality on the 
group $U(N)$ on the configuration $NA + N_F B$. 
This gives us the following quiver diagram

\littlefig{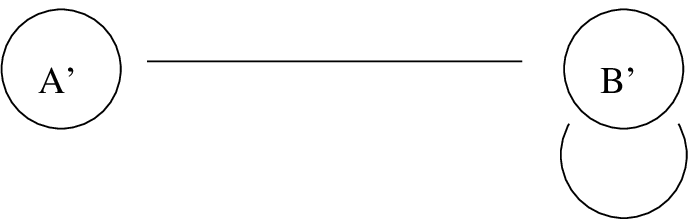}{2.5}

Where the new gauge group is given by $(N_F-N) A' + N_F B'$.  We can now 
take $N<N_F$, or even negative. The advantage is that now we can consider the 
moduli space as being given by the adjoint field $\phi_{B'B'}$, and we also 
have the superpotential
\begin{equation}
\tr(\phi_{B'B'}\phi_{B'A'}\phi_{A'B'})
\end{equation}
If we ignore the $A'$ branes, then we have an ${\cal N}=2$ 
gauge theory on the volume of the branes $B'$, and 
therefore the moduli space is not lifted. However, once we include the branes 
$A'$ it is possible to integrate the quarks for the gauge group $U(N_{A'})$
at generic points in the moduli space for the field $\phi_{B'B'}$, which is 
identified with the meson matrix $M$.

The result of the integration gives an effective potential for the field 
$\phi$ from a disk diagram computation
\begin{equation}\label{eq:su2}
W_{eff} = N_{A'} (S\log S-S)+\tau_{A'} S - S\log\det(M)
\end{equation}
Notice that the only difference between \ref{eq:su1} and \ref{eq:su2}
is the sign of the term that contains the determinant of the meson field. 
This can be argued because in the first case one obtains the term from 
integrating ghosts (a gauge fixing procedure), while in the other case they 
are obtained from integrating matter.

This fits very well with continuing all the results for negative values
of $N$, and considering this as a calculation that was performed in the 
brane category. As discussed in \cite{BD}, Seiberg dualities correspond to 
basis changes for fractional branes, and the natural basis are determined 
by terms in the Kahler potential. In the above example $A' \sim -A$ 
corresponds to the class of the antibrane of $A$ in a different region of the
Khaler moduli where the collection of mutually BPS branes is different.

For this case the Affleck-Dine-Seiberg superpotential looks as follows, 
once we have integrated out $S$
\begin{equation}
W_{eff} \sim  (\det M)^{1/N_{A'}}
\end{equation}
notice that this superpotential grows at infinity in the moduli space. 
However, when we analyze it one eigenvalue of $M$ at a time with 
all others fixed, it grows slower than a polynomial. If we scale 
the meson variables as $M\to t M$, $W_{eff}\sim t^{N_F/N_{A'}}$
which grows slower than $t$ for $N_F<N_{A'}$. In this case the 
effective potential at infinity (for canonical fields) is  given by 
\begin{equation}
|\frac{\partial W}{\partial\phi}|^2 \sim t^{2 N_F/N_{A'} -2}
\end{equation}
goes to zero, and one still has runaway behavior, 
as the total energy will decrease going to infinity.

Notice that in all of these cases there is monodromy of the superpotential
at infinity, so $W_{eff}$ still has to have singularities at 
finite values, because one needs a place where the cut of $W_{eff}$ 
originating at infinity ends. For the most part in this situation 
the
discussion in the previous section about the topological features of moduli 
space goes unchanged: $\partial W/\partial \phi$ still vanishes at the 
boundary and one has monodromies for
$\partial W/\partial \phi$ inside the moduli space, but it does not need to 
vanish in the interior.

Let us now compare this result with the literature.
This type of example has been discussed in \cite{IT}, 
where it was argued that a quantum deformed moduli space was incompatible 
with the F-terms which are produced from adding sources which only appear 
linearly. In the case above, this is the field $\phi_{B'B'}$, 
and the quantum moduli space would correspond to to the bilinears 
$\phi_{B'A'}\phi_{A'B'}$, under special circumstances. Also it was argued 
in various limits that gaugino condensation would produce an effective 
superpotential on the moduli space for the $\phi$.  The difference 
now is that we are not forced to take limits. The matrix model technology 
lets us do a complete analysis on the whole moduli space, independent of 
the couplings.

\section{An example of moduli stabilization}\label{sec:stab}

Let us consider the results of the previous section, particularly the 
last part, where we had a Seiberg-dual version of the ADS 
superpotential. It is interesting to write models where one can produce 
moduli stabilization and not just runaway behavior. The simplest 
such setup is to setup a racetrack scheme: two non-perturbative 
effects compete with each other to stabilize the vacua\cite{Kras}.

The simplest such setup in the considerations we have been making is to take 
a one complex dimensional space with two singularities. At each of the 
singularities place fixed branes which are not allowed to move, but that 
repel the brane with the moduli space from the singularity.

This can be done by introducing two confining gauge groups in the
following quiver diagram

\littlefig{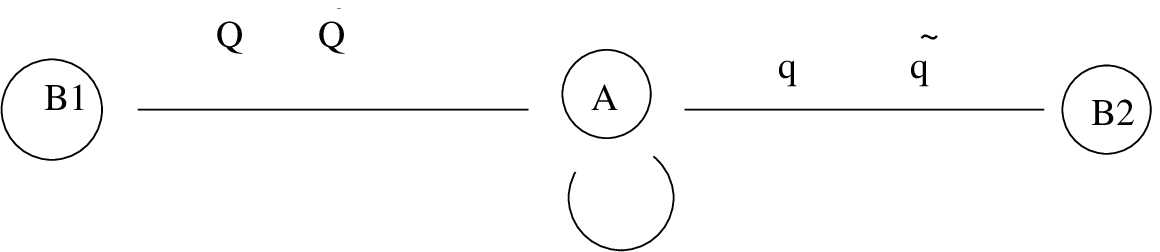}{4}

With a superpotential given by 
\begin{equation}
\phi_A Q\tilde Q + (\phi_A+m)q\tilde q
\end{equation}

At generic points in moduli space for $\phi_A$, both $Q,\tilde Q$ and
$q,\tilde q$ are massive and can be integrated out. The disc diagram
computations produce
an effective potential for $\phi$ and the gaugino condensates for branes 
$B_1,B_2$ given by
\begin{eqnarray}
W_{eff} = N_1 (S_1\log S_1/\Lambda^3-S_1)+\tau_1 S_1 + N_2(S_2\log S_2/\Lambda^3-S_2)\\
+\tau_2 S_2 - S_1\log(\det (\phi/\Lambda)) - S_2\log(\det((\phi+m)/\Lambda)) 
\end{eqnarray}
Again, we can go to the eigenvalue basis for $\phi$ by gauge 
transformations, and we obtain the effective superpotential
for each eigenvalue $\lambda$ as given by
\begin{equation}
-S_1\log(\lambda/m) -S_2\log(\lambda/m+1) 
\end{equation}
which produces a supersymmetric (semiclassical) vacuum at 
\begin{equation}
\lambda = -\frac{S_1}{S_1+S_2}m
\end{equation}

Once the branes are stabilized we see that the vacuum has $U(N_A)$ 
pure gauge symmetry in the IR, and the $SU(N_A)$ will confine producing 
extra non-perturbative effects. 

For the most part we can ignore this possibility if we take just 
$N_A=1$. Even if we have $N_A\neq 1$ one can argue that this can happen at 
very small scales compared to any other scale by judiciously choosing the 
relations between the couplings $\tau_A,\tau_1,\tau_2$ at a very high
scale (the string scale for example). Since we can arrange for branes $A$ to 
become infrared free for $Q$ and $q$ massless, the coupling will run in the 
IR so that it is very weakly coupled at the scale $m$, while at the same 
time we can arrange for
the couplings $\tau_1,\tau_2$ to be of order one at the scale $m$, because 
the matter content is not sufficient to stop the running coupling 
constant from becoming large. This can be arranged even if all of the 
coupling constants $\tau_i$ are perturbative and approximately equal at the 
UV scale.

There are various interesting aspects of the above field theory. It can be 
produced by taking branes on a deformed $A_3$ singularity. This begins
with an 
${\cal N}=2$ SYM theory. One can softly break it to ${\cal N}=1$ by adding 
mass terms for some of the adjoint of the fields, and still keep the 
geometric engineering geometry, reducing the singularity to an $A_1$ 
singularity almost everywhere. These type of geometric constructions
 have been discussed in \cite{CKV,CFIKV}, where they choose
to generically lift the moduli 
space for fractional branes completely.

The $A_3$ singularity has four nodes, and the fourth node is not present in 
the above discussion, so it can be used to make the deformation geometrical.
This produces an example of moduli stabilization in a system of D-branes
with softly broken ${\cal N}=2$ SUSY.

The above example can be phrased also as moduli stabilization by fluxes. Once 
the gauginos of the branes $B_1,B_2$ condense, they give rise to a geometric 
transition and are replaced by fluxes. 
The fractional brane with a moduli space gets 
localized because there is a flux induced superpotential on the brane.

The interesting questions to analyze in this setup are the conditions under 
which we can make $\phi/m$ very large, so that the theory for the group of 
branes $A$ is perturbative so that the fields $Q,\tilde Q, q, \tilde q$  
have a large mass,
 and to ask how much fine tuning  is involved in accomplishing this 
condition.

The simplest case  to set this up is to take the limit $m\to 0$.
 Then one has the two singularities 
in the complex $\phi$ plane coalescing.

The effective superpotential for the field $\phi$ once $S_1,S_2$ are 
integrated out is given by
\begin{equation}
W_{eff} = -( N_1S_1+N_2S_2) = \Lambda^3(A(\phi/\Lambda)^{1/N_1} + B(\phi/\Lambda)^{1/N_2})
\end{equation}
and this expression 
has a saddle point at a non-zero value of $\phi$ so long as 
$N_2\neq N_1$. 
At the minimum of the potential
$S_1=-S_2$, and 
$|\phi|\sim \Lambda\exp^{(N_2\tau_1-N_1\tau_2)/(N_2-N_1)}$. 
Which can be even very large compared to $\Lambda$, without 
much fine tuning on the couplings $\tau_i$.

One can ask what the value of $\phi$ is for a weakly coupled setup
for $\tau_1= \tau_2=\tau$ and large (lets say at the string scale).
Then one easily sees that 
$|\phi|\sim \Lambda|\exp\tau|$.

For our purposes it is better to 
keep $m$ in the discussion since we can 
also find moduli stabilization in 
the case $N_1=N_2$.
With just one fractional brane, the saddle point equations for $S_1,S_2$
give the following result
\begin{eqnarray}
S_1 &=& \Lambda^3 e^{-\tau_1/N_1 + 2\pi i k_1/N_1} (\phi/\Lambda)^{1/N_1}\\
S_2 &=& \Lambda^3 e^{-\tau_2/N_2+2\pi i k_2/N_2}((\phi+m)/\Lambda)^{1/N_2}
\end{eqnarray}
where $k_1,k_2$ are integers which choose the vacua of the two 
$U(N_i)$ gauge groups. Since generically $S_1,S_2$ 
have different dependence on
$\phi$ it is possible to give vevs to $\phi$ so that $S_1\sim -S_2$
and if $\phi$ is large with respect to $m$ then it will look like the
previous calculation. It 
is still difficult to solve for $\phi/m = -S_1/(S_1+S_2)$. In general 
this will give rise to a polynomial problem for $\phi$ of very high degree
($N_1N_2$ in most cases).

Notice that one can also take phases where $S_1,S_2$ are not oriented 
opposite to each other, and these give vevs to $\phi$ which are 
of order $m$. If $m<<\Lambda$ one can not trust the Kahler potential for the
field $\phi$ and one can not estimate the mass for the field $\phi$
reliably.

When $\phi/m$ is large we can ignore the $m$ dependence on the 
$S_i$, and the theory reduces to $m\sim 0$, so we can 
trust the effective lagrangian for $\phi$, as it ends up with 
a large vev compared to 
$\Lambda$. From here one can estimate the mass of the field $\phi$
from the effective action 
\begin{equation}
S_{eff} \sim \int d^4\theta \frac{1}{ g^2 }
\phi^\dagger\phi+\int d^2\theta W_{eff}(\phi)
\end{equation}
and where $g$ could be the associated ${\cal N}=2$ coupling.
The mass for $\phi$ is then of order 
\begin{equation}
g^2 \frac{\partial^2 W_{eff}}{\partial\phi^2} 
\sim g^2 \Lambda^3/\phi^2
\end{equation}
which can be very small if $g$ is small and 
perturbative. Having $\phi$ perturbative is more or less equivalent to 
requiring that the dilaton is fixed at a point 
where perturbation theory is valid 
in other setups.

Notice that the mechanism that chooses the large vev of $\phi$ depends on 
choosing the right vacuum among $N_1N_2$ of them. Without any fine tuning it 
was possible to find vacua with large vevs of $\phi$.
This is a discrete choice
that has the same structure as the one
described in \cite{BP}, where there is one 
choice
among  
a large set of discrete fluxes that can balance the cosmological 
constant (this was called the discretuum of vacua: among the very 
many possible 
solutions with generic behavior,  there is a good chance of finding one 
good vacuum). 

Also for this one vacuum the
vev of $W_{eff}$ is supressed with respect to other vacua, as it is given by 
$-N_1S_1-N_2S_2$. If this were a situation where gravity has not been 
decoupled 
because of the infinite volume CY, then at these values of $S_i$ the 
contributions of 
$S_1$ and $S_2$ would work against each other making the cosmological constant 
small (but negative)  at the same time.

In this example, on top of guaranteeing that we have one light particle
that sits at a perturbative value, we also have at the same time tuned the 
vev of the superpotential to be relatively small compared to other 
vacua, 
without requiring any additional 
tuning. In this sense we have found an example where two effects which are 
desirable for model building are correlated.

\section{Conclusion}

In this paper we have presented various constructions where 
D-brane moduli spaces
are lifted by non-perturbative quantum corrections, sometimes producing 
supersymmetric vacua where the moduli fields are light and have a large 
vev. 

These 
moduli spaces are 
usually for a D-brane which is located at a curve of singularities in a 
Calabi-Yau geometry. Because we studied non-compact Calabi-Yau geometries it
was possible to reduce the problem to ordinary supersymmetric field theory, 
and one 
could treat the moduli stabilization mechanism exactly by using matrix 
model techniques.
These situations are fairly simple to engineer and are 
a variation on racetrack schemes to stabilize the moduli 
fields.

It would be interesting if these models could be extended to a setup where
gravity is dynamical in four dimension and where one can also understand
supersymmetry 
breaking in a controllable manner.

\section*{Acknowledgements}

I would like to thank M. Douglas, I. Klebanov,  J. Maldacena, N. Seiberg
for many useful discussions. Research supported in par by DOE grant
DE-FG02-90ER40452.

\appendix

\section{Derivation of the deformation of the $\BC^3/\BZ_2\times\BZ_2$ 
geometry}\label{sec:appendix}

The quiver diagram for the $\BC^3/\BZ_2\times \BZ_2$ geometry is given 
below.

\littlefig{quiver1.eps}{2}

The superpotential of the theory is given by the following expression
\begin{eqnarray}
W &=& \tr( \phi_{AB}\phi_{BD}\phi_{DA}-\phi_{AB}\phi_{BC}\phi_{CA}\nonumber\\
&&+\phi_{BA}\phi_{AC}\phi_{CB}-\phi_{BA}\phi_{AD}\phi_{DB}\label{eq:WClas}\\
&&+\phi_{CD}\phi_{DB}\phi_{BC}-\phi_{CD}\phi_{DA}\phi_{AC}\nonumber\\
&&+\phi_{DC}\phi_{CA}\phi_{AD}-\phi_{DC}\phi_{CB}\phi_{BD})\nonumber
\end{eqnarray}

We want to choose the theory with brane content $NA + (A+B+C+D)$, this is, 
one brane in the bulk in the presence of fractional branes at the 
singularity. To calculate the moduli space of the brane in the bulk, and 
therefore the complex structure of the deformation, we need to split the
$U(N+1)$ indices of $\phi_{XA}$ and $\phi_{AX}$ into the $U(N)$ singlet and 
the part in the fundamental. The singlet is going to be part of the moduli 
space of vacua, so it will not be integrated out, while the 
matter in the fundamental will be massive and can be integrated out.
This manner of calculating has been described in detail in the
papers \cite{Bqu,Bhol}. 

Here we will just include the partial gaugino condensate $S$ for the 
$U(N)$ unbroken gauge group. The masses for the fields which are charged under 
the $U(N)$  gauge field are given by the $3\times3$ matrix
\begin{equation}
M = \begin{pmatrix}0&-\phi_{BC}&\phi_{BD}\\
\phi_{CB}&0&-\phi_{CD}\\
-\phi_{DB}&\phi_{CD}&0
\end{pmatrix}
\end{equation}
Before we just include the determinant of the mass term, we need to use a 
gauge fixing procedure to get the $U(N)$ gauge group inside the $U(N+1)$ 
group, this gives a contribution from ghosts in the matrix model 
\cite{DGKV}. In situations with adjoints this is the contribution from the 
Vandermonde determinant. Here, we get instead, by gauge fixing 
$\phi_{AB}$ the following effective superpotential:
\begin{equation}\label{eq:weff}
W_{eff} = -S\log(\phi_{AB}\phi_{BA})
\end{equation}
which makes us take the color component of $\phi_{AB}$ and set it to zero.
The contribution from the mass term of the field $\phi_{AC}, \phi_{AD},
\phi_{CA},\phi_{DA}$ then gives us
\begin{equation}\label{eq:weff2}
W_{eff2} = S\log(\phi_{CD}\phi_{DC})
\end{equation}
The total effective superpotential for the moduli fields 
is the sum of equations 
\ref{eq:WClas} \ref{eq:weff} and \ref{eq:weff2}.

Now, let us consider the gauge invariant variables 
$u=\phi_{BA}\phi_{AB}, v = \phi_{BC}\phi_{CB}, w = \phi_{BD}\phi_{DB}$
and $t=\phi_{BC}\phi_{CA}\phi_{AB}$.
If we square $t$ we obtain
\begin{equation}
t^2 = \phi_{BC}\phi_{CA}\phi_{AB}\phi_{BC}\phi_{CA}\phi_{AB}
\end{equation}
Now we can use the deformed (by $S$)
F-term equations of motion for the moduli fields to change the subscripts 
as follows
\begin{eqnarray}
t^2 &=& \phi_{BC}\phi_{CA}\phi_{AD}\phi_{DC}\phi_{CA}\phi_{AB}\\
&=&\phi_{BC}[\phi_{CB}\phi_{BD}-S/\phi_{DC}]\phi_{DC}\phi_{CA}\phi_{AB}\\
&=&S \phi_{BC}\phi_{CA}\phi_{AB}+\phi_{BC}\phi_{CB}\phi_{BD}\phi_{DB}
\phi_{BA}\phi_{AB}\\
&=& S t + uvw
\end{eqnarray}
After a linear change of variables in $t$ we can turn this expression into
the form
\begin{equation}
uvw = t^2 +c
\end{equation}
which is exactly what was expected due to holomorphy arguments in equation
\ref{eq:z2z2def}.


\begin{thebibliography}{99}
\bibitem{ADS}
I.~Affleck, M.~Dine and N.~Seiberg,
``Dynamical Supersymmetry Breaking In Supersymmetric QCD,''
Nucl.\ Phys.\ B {\bf 241}, 493 (1984).
\bibitem{BR}
I.~Bena and R.~Roiban,
``Exact superpotentials in N = 1 theories with flavor and their matrix  model formulation,''
Phys.\ Lett.\ B {\bf 555}, 117 (2003)
[arXiv:hep-th/0211075].


\bibitem{Bqu}
D.~Berenstein,
``Quantum moduli spaces from matrix models,''
Phys.\ Lett.\ B {\bf 552}, 255 (2003)
[arXiv:hep-th/0210183].


\bibitem{Bhol}
D.~Berenstein,
``Solving matrix models using holomorphy,''
arXiv:hep-th/0303033.


\bibitem{BD}
D.~Berenstein and M.~R.~Douglas,
``Seiberg duality for quiver gauge theories,''
arXiv:hep-th/0207027.



\bibitem{BP}
R.~Bousso and J.~Polchinski,
``Quantization of four-form fluxes and dynamical neutralization of the  cosmological constant,''
JHEP {\bf 0006}, 006 (2000)
[arXiv:hep-th/0004134].

\bibitem{CKV}
F.~Cachazo, S.~Katz and C.~Vafa,
``Geometric transitions and N = 1 quiver theories,''
arXiv:hep-th/0108120.

\bibitem{CFIKV}
F.~Cachazo, B.~Fiol, K.~A.~Intriligator, S.~Katz and C.~Vafa,
``A geometric unification of dualities,''
Nucl.\ Phys.\ B {\bf 628}, 3 (2002)
[arXiv:hep-th/0110028].

\bibitem{CDSW}
F.~Cachazo, M.~R.~Douglas, N.~Seiberg and E.~Witten,
``Chiral rings and anomalies in supersymmetric gauge theory,''
JHEP {\bf 0212}, 071 (2002)
[arXiv:hep-th/0211170].

\bibitem{DJ}
Y.~Demasure and R.~A.~Janik,
``Effective matter superpotentials from Wishart random matrices,''
Phys.\ Lett.\ B {\bf 553}, 105 (2003)
[arXiv:hep-th/0211082].


\bibitem{DGLVZ}
R.~Dijkgraaf, M.~T.~Grisaru, C.~S.~Lam, C.~Vafa and D.~Zanon,
``Perturbative computation of glueball superpotentials,''
arXiv:hep-th/0211017.

\bibitem{DGKV}
R.~Dijkgraaf, S.~Gukov, V.~A.~Kazakov and C.~Vafa,
``Perturbative analysis of gauged matrix models,''
arXiv:hep-th/0210238.

\bibitem{DV}
R.~Dijkgraaf and C.~Vafa,
``A perturbative window into non-perturbative physics,''
arXiv:hep-th/0208048.

\bibitem{DS}
M.~Dine and Y.~Shirman,
``Remarks on the racetrack scheme,''
Phys.\ Rev.\ D {\bf 63}, 046005 (2001)
[arXiv:hep-th/9906246].


\bibitem{Dou}
M.~R.~Douglas,
``The statistics of string/M theory vacua,''
arXiv:hep-th/0303194.

\bibitem{IT}
K.~A.~Intriligator and S.~Thomas,
``Dynamical Supersymmetry Breaking on Quantum Moduli Spaces,''
Nucl.\ Phys.\ B {\bf 473}, 121 (1996)
[arXiv:hep-th/9603158].


\bibitem{GV}
R.~Gopakumar and C.~Vafa,
``On the gauge theory/geometry correspondence,''
Adv.\ Theor.\ Math.\ Phys.\  {\bf 3}, 1415 (1999)
[arXiv:hep-th/9811131].

\bibitem{KKLT}
S.~Kachru, R.~Kallosh, A.~Linde and S.~P.~Trivedi,
``De Sitter vacua in string theory,''
arXiv:hep-th/0301240.

\bibitem{KL}
V.~Kaplunovsky and J.~Louis,
``Phenomenological aspects of F-theory,''
Phys.\ Lett.\ B {\bf 417}, 45 (1998)
[arXiv:hep-th/9708049].

\bibitem{KS}
I.~R.~Klebanov and M.~J.~Strassler,
``Supergravity and a confining gauge theory: Duality cascades and  chiSB-resolution of naked singularities,''
JHEP {\bf 0008}, 052 (2000)
[arXiv:hep-th/0007191].



\bibitem{Kras}
N.~V.~Krasnikov,
``On Supersymmetry Breaking In Superstring Theories,''
Phys.\ Lett.\ B {\bf 193}, 37 (1987).


\bibitem{Ohta}
K.~Ohta,
``Exact mesonic vacua from matrix models,''
JHEP {\bf 0302}, 057 (2003)
[arXiv:hep-th/0212025].

\bibitem{S}
L.~Susskind,
``The anthropic landscape of string theory,''
arXiv:hep-th/0302219.

\bibitem{Sexact}
N.~Seiberg,
``Exact results on the space of vacua of four-dimensional 
SUSY gauge theories,''
Phys.\ Rev.\ D {\bf 49}, 6857 (1994)
[arXiv:hep-th/9402044].

\bibitem{Shol}
N.~Seiberg,
``The Power of holomorphy: Exact results in 4-D SUSY field theories,''
arXiv:hep-th/9408013.

\bibitem{Snal}
N.~Seiberg,
``Naturalness versus supersymmetric nonrenormalization theorems,''
Phys.\ Lett.\ B {\bf 318}, 469 (1993)
[arXiv:hep-ph/9309335].

\bibitem{SW}
N.~Seiberg and E.~Witten,
``Electric - magnetic duality, monopole condensation, and confinement in N=2 supersymmetric Yang-Mills theory,''
Nucl.\ Phys.\ B {\bf 426}, 19 (1994)
[Erratum-ibid.\ B {\bf 430}, 485 (1994)]
[arXiv:hep-th/9407087].


\bibitem{SS}
Y.~Shadmi and Y.~Shirman,
``Dynamical supersymmetry breaking,''
Rev.\ Mod.\ Phys.\  {\bf 72}, 25 (2000)
[arXiv:hep-th/9907225].

\bibitem{Wnp}
E.~Witten,
``Non-Perturbative Superpotentials In String Theory,''
Nucl.\ Phys.\ B {\bf 474}, 343 (1996)
[arXiv:hep-th/9604030].



\end{thebibliography}
\end{document}